\documentstyle[prl,aps,epsf,multicol,tabularx]{revtex}

\begin{document}
\draft

\title{Anomalous quantum reflection of Bose-Einstein condensates from a silicon surface: the role of dynamical excitations}

\author{R.G. Scott$^1$, A.M. Martin$^{2}$, T.M. Fromhold$^{1}$ and F.W. Sheard$^{1}$\\\textit{$^{1}$School of Physics and Astronomy, University of Nottingham,}\\\textit{Nottingham NG7 2RD, United Kingdom}\\\textit{$^{2}$School of Physics, University of Melbourne, Parkville, Vic. 3010, Australia}}

\date{\today}
 
\maketitle
\begin{abstract}
We investigate the effect of inter-atomic interactions on the quantum-mechanical reflection of Bose-Einstein condensates from regions of rapid potential variation. The reflection process depends critically on the density and incident velocity of the condensate. For low densities and high velocities, the atom cloud has almost the same form before and after reflection. Conversely, at high densities and low velocities, the reflection process generates solitons and vortex rings that fragment the condensate. We show that this fragmentation can explain the anomalously low reflection probabilities recently measured for low-velocity condensates  incident on a silicon surface.
\end{abstract}

\begin{multicols}{2}
Advances in atom cooling and trapping have enabled the exploration of diverse quantum transport and interference effects for both individual atoms~\cite{dahan} and Bose-Einstein condensates (BECs)~\cite{pethick,morsch,andrews,mewes,chikkatur,bloch,myprl,mypra,janne,wu,fallani,tom,shimizu,choi,arimondo,zheng}. In BECs, inter-atomic interactions can cause complex collective excitations of the atom cloud, such as soliton and vortex formation~\cite{myprl,mypra,janne}, and dynamical and Landau instabilities~\cite{wu,fallani,choi,arimondo,zheng,raman}, which strongly influence the BEC's center-of-mass motion. There is great current interest in using condensed matter systems to provide new tools for manipulating ultra-cold atoms, for example micron-scale current-carrying wires on a semiconductor surface~\cite{hansel}.  Understanding and controlling interactions between the atoms and the surface is crucial for the development of these ``atom chips''. Recent experimental studies of Na-atom BECs bouncing off a Si surface provide a powerful probe of such interactions~\cite{tom}. When the approach velocity, $v_{x}$, is low, the rapid variation of the surface potential causes quantum-mechanical reflection (QR) of the atom cloud. 
Although quantum-mechanical calculations predict that the reflection probability for a \textit{single} atom will increase monotonically as $v_{x}$ decreases~\cite{tom,shimizu}, the measured reflection probability for the BEC \textit{decreases} with decreasing $v_{x}$ below $\sim 1.7$ mm s$^{-1}$: an interesting observation that is not yet understood~\cite{tom}.

In this Letter, we show that inter-atomic interactions can have a pronounced effect on the QR of BECs from Si surfaces, consistent with the anomalously low reflection probabilities observed in experiment~\cite{tom}. Our results demonstrate that the underlying physics is a \textit{generic} feature of the reflection of BECs from regions of rapid potential variation, attractive or repulsive.  
As the BEC approaches the surface, its density profile is modulated by the standing wave formed by superposition of incident and reflected matter waves.
Due to the inter-atomic interactions in a BEC, this modulation can have a dramatic effect on the reflection process. In particular, for high density BECs and low incident velocities, it triggers the formation of dynamical excitations (solitons and vortex rings) that fragment the atom cloud. 
Our analysis shows that although this fragmentation has no \textit{intrinsic} effect on the reflection probability, it disperses the atoms and can therefore produce an \textit{apparent} reduction in the reflection probability at low velocities, as observed in experiment~\cite{tom}. Our calculations enable us to identify regimes in which the fragmentation of the atom cloud that accompanies low-speed QR should be either strongly enhanced or suppressed, and thereby propose new experiments to test our interpretation of the reflection process.

\begin{figure}
\narrowtext
\vspace{-0.1cm}
\epsfxsize=4cm \centerline{\epsffile{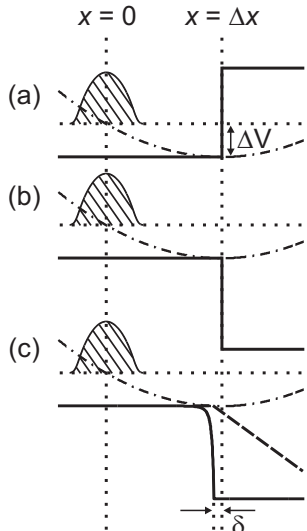}} \vspace{-0.0cm}
\caption{Solid curves: schematic representations of $V_{\mbox{I}}\left(x\right)$ (a), $V_{\mbox{II}}\left(x\right)$ (b), and $V_{\mbox{III}}\left(x\right)$ (c). Dot-dashed curves: the potential energy of the harmonic trap, shown as a function of $x$, immediately after the trap displacement. Shaded areas in (a)-(c) represent the initial atom density profile $\left|\Psi\left(x,0,0\right)\right|^{2}$. Dashed line in (c) shows the imaginary potential used to model the adsorption of {N}a atoms on the {S}i surface.}
\label{fig1}
\end{figure}

We consider BECs containing $N$ $^{23}${N}a atoms in a cylindrically symmetrical harmonic trap. First we take parameters from recent experiments~\cite{tom}: $N = 3 \times 10^{5}$ and longitudinal (radial) trap frequencies $\omega_{x} \left(\omega_{r}\right) = 2\pi \times 3.3 \left(2\pi \times 4.5\right)$ rad s$^{-1}$, which produce an atom cloud of equilibrium peak density $n_{0}=2.2 \times 10^{12}$ cm$^{-3}$. 
At time $t=0$ ms, we suddenly displace the harmonic trap by a distance $\Delta x$ along the  $x$-direction~\cite{tom}, and hence accelerate the BEC towards a region of rapid potential energy variation at $x = \Delta x$. We consider three different potential profiles: potential I [$V_{\mbox{I}}\left(x\right)$, solid curve in Fig.~\ref{fig1}(a)] is a sharp potential step of height $V_{s}=10^{-30}$ J; potential II [$V_{\mbox{II}}\left(x\right)$, solid curve in Fig.~\ref{fig1}(b)] is an abrupt potential drop of depth $V_{s}$; potential III [$V_{\mbox{III}}\left(x\right)$, solid curve in Fig.~\ref{fig1}(c)] is a model of the Si surface, whose form is specified later. The model potentials I and II enable us to identify the key physical processes that occur when a BEC undergoes reflection, and thereby develop the understanding required to interpret experimental studies of BECs reflecting from the more complicated Si surface potential. Following the trap displacement, the total potential energy of each {N}a atom (mass $m$) in the {BEC} is $V_{T}\left(x,r\right)= V_{\mbox{step}}\left(x\right)+ \frac{1}{2}m[\omega_{x}^{2}\left(x-\Delta x\right)^{2}+\omega_{r}^{2}r^{2}]$, where $r$ is the radial coordinate and $V_{\mbox{step}}\left(x\right)$ is $V_{\mbox{I}}\left(x\right)$, $V_{\mbox{II}}\left(x\right)$, or $V_{\mbox{III}}\left(x\right)$.

\begin{figure}
\narrowtext
\vspace{-0.1cm}
\epsfxsize=7.0cm \centerline{\epsffile{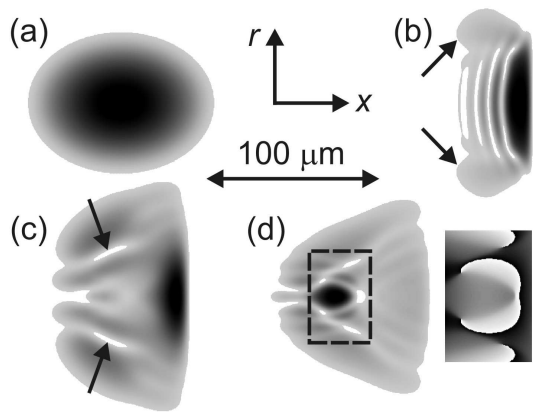}} \vspace{0.1cm}
\caption{Grey-scale plots of atom density (black = high) in the $x-r$ plane (axes inset) for a {BEC} with $n_{0}=2.2 \times 10^{12}$ cm$^{-3}$, $\omega_{x}$ $\left(\omega_{r}\right)$ $= 2\pi \times 3.3$ $\left(2\pi \times 4.5\right)$ rad s$^{-1}$ and $v_{x}\approx1.2$ mm s$^{-1}$, reflecting from potential I, shown at $t = 0$ ms (a), $90$ ms (b), $122$ ms (c) and $143$ ms (d). The phase of the BEC wavefunction within the dashed box in (d) is shown in the grey-scale plot (white = 0, black = $2\pi$) to the right of the density profile. Horizontal bar shows scale.}
\label{fig2}
\vspace{-0.1cm}
\end{figure}

We determine the dynamics of the BEC in three dimensions by using the Crank-Nicolson~\cite{choi} method to solve the time-dependent Gross-Pitaevskii equation~\cite{myprl,mypra}
\begin{equation}
i\hbar \frac{\partial \Psi }{\partial t} =
\left[
-\frac{\hbar^2}{2m}\nabla^2 + V_T\left(x,r\right) +
\frac{4 \pi \hbar^2 a}{m} \left| \Psi \right|^{2}
\right] \Psi
\label{eq:gp}
\end{equation}
where $\nabla^2$ is the Laplacian in cylindrical coordinates, $a=2.9$ nm is the s-wave scattering length and $\Psi\left(x,r,t\right)$ is the axially symmetrical condensate wavefunction at time $t$, normalized such that $\left|\Psi\right|^{2}$ is the number of atoms per unit volume. The equilibrium density profile of the BEC in the harmonic trap is shown in Fig.~\ref{fig2}(a).

\begin{figure}
\narrowtext
\vspace{-0.3cm}
\epsfxsize=7.5cm \centerline{\epsffile{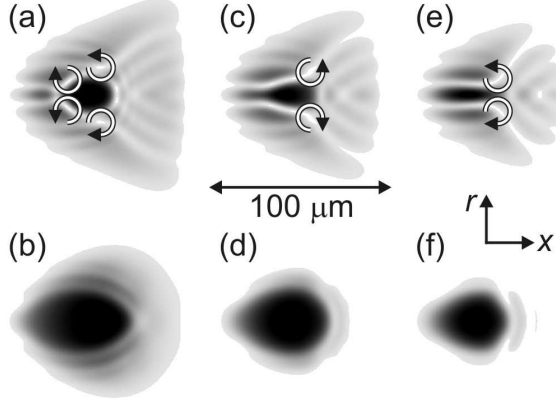}} \vspace{0.0cm}
\caption{Grey-scale plots of atom density (black = high) in the $x-r$ plane (axes inset) for a {BEC} with with $n_{0}=2.2 \times 10^{12}$ cm$^{-3}$, $\omega_{x}$ $\left(\omega_{r}\right)$ $= 2\pi \times 3.3$ $\left(2\pi \times 4.5\right)$ rad s$^{-1}$ and $v_{x}\approx1.2$ mm s$^{-1}$ (a), $2.1$ mm s$^{-1}$ (b), reflecting from potential I, shown at $t = 150$ ms. (c) \& (d) Same as (a) \& (b), but for potential II. (e) \& (f) Same as (a) \& (b), but for potential III. Horizontal bar shows scale. Arrows in (a), (c) and (e) indicate the direction of quantized circulation around vortex cores.}
\vspace{-0.1cm}
\label{fig3}
\end{figure}

We now explore the dynamics of the BEC in potential I. When the trap is displaced at $t=0$ ms, the initial potential energy of the BEC is increased by $\Delta V \approx \frac{1}{2}m\omega_{x}^{2} \Delta x^{2}$, as shown in Fig.~\ref{fig1}(a), causing the atoms to be incident on the potential step with a velocity $v_{x} \approx\omega_{x} \Delta x$. For the displacements considered here, $\Delta V << V_{s}$. Hence, in a classical picture, we expect the {BEC} to be completely reflected by the step. Figure~\ref{fig2} shows images of the {BEC} undergoing reflection from potential I after a trap displacement of 60 $\mu$m, for which $v_{x}\approx1.2$ mm s$^{-1}$. At $t=90$ ms [Fig.~\ref{fig2}(b)], approximately half the atoms have reached the barrier, and a standing wave has formed as a result of the superposition of the incident and reflected waves, causing periodic modulations in the density profile. Due to the inter-atomic interactions, the high density at the peaks of the standing wave causes atoms to be pushed out from the axis of cylindrical symmetry (the $x$-axis), thus transferring momentum into the transverse direction. These atoms appear as the arrowed jet-like ``lobes'' at the top and bottom of Fig.~\ref{fig2}(b). The lobes ride up the walls of the harmonic trap and are then reflected back towards the $x$-axis. This causes a cylindrically symmetrical soliton to form, which appears as the two white stripes (arrowed) in Fig.~\ref{fig2}(c). The soliton then decays via the snake instability~\cite{dutton} into two vortex rings centered on $r=0$, which appear as pairs of density nodes within the dashed box in Fig.~\ref{fig2}(d). The phase of the {BEC} wavefunction within the dashed box in Fig.~\ref{fig2}(d) is shown to the right of the density profile. The figure reveals a $2\pi$ phase change around each density node, indicating quantized circulation. At the end of the oscillation, the atom cloud still contains vortex rings, and has a fragmented appearance [Fig.~\ref{fig3}(a)]. Similar dynamics have been observed for two-dimensional {BEC}s reflecting from a circular hard wall~\cite{janne}.

For disruption to occur, the time taken for lobes to form, $t_{L}\approx l_{r}/v_{s}$, where $l_{r}$ is the BEC's radial width and $v_{s} \approx \sqrt{h^{2}n_{0}a/ \pi m^{2}}$ is the average speed of sound ~\cite{pethick}, must be less than the reflection time $t_{R}\approx l_{x}/v_{x}$, where $l_{x}$ is the BEC's longitudinal width. We therefore estimate that reflection will \emph{not} significantly disrupt the BEC when 
\begin{equation}
v_{x} \gtrsim v_{s}\left(l_{x}/l_{r}\right),
\label{eq:condition}
\end{equation} 
or, equivalently, when $\Delta x \gtrsim \left(v_{s}l_{x}\right)/\left(l_{r}\omega_{x}\right)$.
Our numerical simulations confirm this~\cite{note9}: for a large $v_{x}\approx2.1$ mm s$^{-1}$ ($\Delta x=100$ $\mu$m), the reflected atom cloud at $t=150$ ms [Fig.~\ref{fig3}(b)] contains no excitations and has a smooth density profile, similar to the original {BEC} ground state [Fig.~\ref{fig2}(a)]. 
Similar behavior has been reported for BECs on Bragg reflection in optical lattices~\cite{mypra,arimondo,zheng,note5}.


We now consider QR of a BEC from the abrupt potential drop (potential II) shown in Fig.~\ref{fig1}(b). 
The reflected atom clouds at $t=150$ ms for $v_{x}\approx1.2$ and $2.1$ mm s$^{-1}$ are shown in Figs.~\ref{fig3}(c) and (d) respectively. 
They are qualitatively remarkably similar to the equivalent images for potential I [Figs.~\ref{fig3}(a) and (b)], except that they are slightly smaller because fewer atoms are reflected. When $v_{x}\approx1.2$ mm s$^{-1}$, the reflected atom cloud has a fragmented appearance and contains a vortex ring, enclosed by the arrows in Fig.~\ref{fig3}(c). By contrast, when $v_{x}\approx2.1$ mm s$^{-1}$ the {BEC} contains no excitations and has a smooth density profile [Fig.~\ref{fig3}(d)]. This is because, just as for potential I, when $v_{x}\approx2.1$ mm s$^{-1}$ the reflection process is too fast for side lobes to form in response to the standing wave. 

Having studied two simple abrupt potential steps, we now investigate QR from a Si surface~\cite{tom} at $x=\Delta x$, which we model by potential III [$V_{\mbox{III}}\left(x\right)$, solid curve in Fig.~\ref{fig1}(c)]. Potential III is based on the Casimir-Polder potential $V_{CP}\left(x'\right) = -C_{4}/x'^{3}\left(x'+3\lambda_{a}/2\pi^{2}\right)$, where $C_{4}=9.1\times10^{-56}$ J m$^{4}$, $\lambda_{a}=590$ nm is the effective atomic transition wavelength~\cite{casimir}, and $x'=x-\Delta x$ is the distance from the surface. 
For $x< \Delta x - \delta$, where $\delta = 0.15$ $\mu$m is a small offset from the surface [Fig.~\ref{fig1}(c)], we set $V_{\mbox{III}}\left(x\right)=V_{CP}\left(x'\right)$. However, since $V_{CP}\rightarrow -\infty$ as $x\rightarrow \Delta x$, for $x\geq\Delta x-\delta$ we set $V_{\mbox{III}}\left(x\right)=V_{CP}\left(\delta\right)+i\left(x-\Delta x+\delta \right) V_{\mbox{im}}$, where $V_{\mbox{im}}=1.6\times10^{-26}$ J m$^{-1}$. No atoms are reflected when $x\geq\Delta x-\delta$ because the real part of $V_{\mbox{III}}\left(x\right)$ is constant. The imaginary part of $V_{\mbox{III}}\left(x\right)$ [dashed curve in Fig.~\ref{fig1}(c)] models the adsorption of Na atoms by the Si surface~\cite{note3}. We expect that the {BEC} will undergo QR when 
$\left|d\lambda/dx\right|\gtrsim 1$~\cite{shimizu}, where $\lambda$ is the local de Broglie wavelength. For $v_{x}= 1$ mm s$^{-1}$, the atoms are most likely to reflect $\sim 1 \mu$m from the surface, where $\left|d\lambda/dx\right|$ is maximal~\cite{note}.

Figures \ref{fig3}(e) and (f) show the reflected atom clouds at $t=150$ ms for $v_{x}\approx1.2$ and $2.1$ mm s$^{-1}$ respectively. Although they are smaller than the corresponding images for potentials I and II because fewer atoms are reflected, the qualitative behavior of the {BEC} is the same. When $v_{x}\approx1.2$ mm s$^{-1}$, the atom cloud is disrupted and contains a vortex ring [Fig.~\ref{fig3}(e)], whereas for $v_{x}\approx2.1$ mm s$^{-1}$ the {BEC} has a smooth density profile and contains no excitations [Fig.~\ref{fig3}(f)]. This is consistent with the experiments of Pasquini et al.~\cite{tom}, who observed ``excited and sometimes fragmented'' atom clouds following the QR of {BEC}s incident on a Si surface at low $v_{x}$. 

\begin{figure}
\narrowtext
\vspace{-0.3cm}
\epsfxsize=5cm \centerline{\epsffile{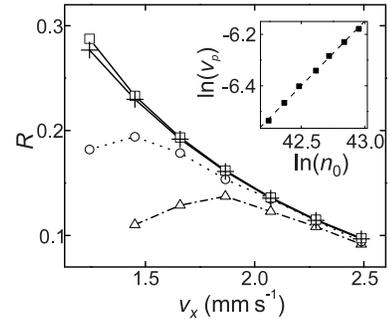}} \vspace{0.1cm}
\caption{Solid curves: $R\left(v_{x}\right)$ for potential III, calculated from the entire reflected atom cloud with (squares) and without (crosses) inter-atomic interactions. Dotted curve with circles: $R\left(v_{x}\right)$ calculated with inter-atomic interactions, but omitting regions of the atom cloud where $\left|\Psi\right|^{2}\leq 0.25 n_{0}$. For the solid and dotted curves, $n_{0}=2.2 \times 10^{12}$ cm$^{-3}$ and $\omega_{x}$ $\left(\omega_{r}\right)$ $= 2\pi \times 3.3$ $\left(2\pi \times 4.5\right)$ rad s$^{-1}$. Dot-dashed curve with triangles: as dotted curve with circles, but for a BEC with $n_{0}=3.6 \times 10^{12}$ cm$^{-3}$. Inset: data points (squares) show $\ln\left(v_{p}\right)$ vs. $\ln\left(n_{0}\right)$, where $v_{p}$ is the position of the peak in $R\left(v_{x}\right)$ curves calculated omitting regions where $\left|\Psi\right|^{2}\leq 0.25 n_{0}$, 
and dashed line is least squares fit.}
\label{fig4}
\end{figure}

To quantify how the QR process depends on $v_{x}$, and to relate our simulations to experiment~\cite{tom}, we made numerical calculations of the probability, $R$, that each atom in the {BEC} will quantum reflect from the Si surface potential. The solid curves in Fig.~\ref{fig4} show $R$ versus $v_{x}$ calculated for atom clouds with (squares) and without (crosses) inter-atomic interactions. In each case, $R$ increases with decreasing $v_{x}$. 
Similar behavior was reported for \emph{single} atoms in Refs.~\cite{tom,shimizu,friedrich}. 

A key feature of Fig.~\ref{fig4} is that the $R\left(v_{x}\right)$ curves calculated with and without inter-atomic interactions are nearly identical~\cite{note4}, indicating that the interactions have little \textit{intrinsic} effect on the reflection probability~\cite{note2}. Consequently, inter-atomic interactions do not \textit{immediately} explain why experimentally measured $R$ values \textit{decrease} with decreasing $v_{x} \lesssim 1.7$ mm s$^{-1}$~\cite{tom}. However, a possible explanation of this anomalous behavior does emerge when we consider how the interactions affect the structure of the BEC at low $v_{x}$. In the absence of interactions, QR does \emph{not} change the form of the atom cloud. By contrast, interactions cause disruption and fragmentation of the atom cloud when $v_{x} \lesssim 1.7$ mm s$^{-1}$ ($\Delta x \lesssim 80$ $\mu$m): precisely the regime where anomalously low $R$ values are observed in experiment~\cite{tom}. Severe disruption of the cloud can make detection of atoms difficult in low density regions of the BEC~\cite{morschemail}, facilitate thermal excitation, or even eject atoms from the trap~\cite{fallani}, thus accounting for the anomalously low $R$ values observed in experiment. 
To simulate the effect of such losses, we recalculated $R$ omitting the contribution from atoms in low density regions of the BEC for which $\left|\Psi\right|^{2}\leq n_{c}$, where $n_{c}$ is a threshold below which detection of atoms can be difficult~\cite{morschemail}.
The dotted curve with circles in Fig.~\ref{fig4} shows $R\left(v_{x}\right)$ calculated for $n_{c}=0.25$ $n_{0}$. In this case, as in experiment~\cite{tom}, $R$ decreases with decreasing $v_{x}<v_{p} \approx 1.7$ mm s$^{-1}$; in the regime where fragmentation of the atom cloud produces large areas of low density~\cite{note6}. 

\begin{figure}
\narrowtext
\vspace{-0.3cm}
\epsfxsize=7cm \centerline{\epsffile{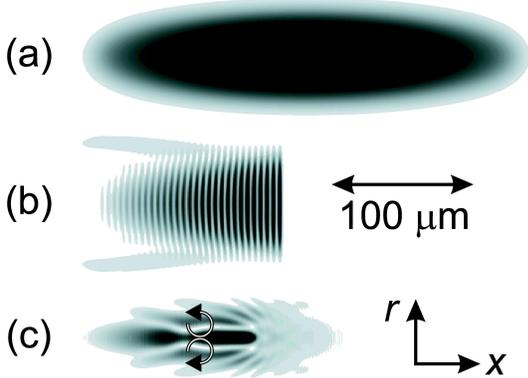}} \vspace{0.0cm}
\caption{Grey-scale plots of atom density (black = high) in the $x-r$ plane (axes inset) for a {BEC} with $n_{0}=2.2 \times 10^{12}$ cm$^{-3}$, $\omega_{x}$ $\left(\omega_{r}\right)$ $= 2\pi$ $\left(2\pi \times 4.5\right)$ rad s$^{-1}$ and $v_{x}\approx2.1$ mm s$^{-1}$, reflecting from potential III, shown at $t = 0$ ms (a), 270 ms (b), and 460 ms (c). Horizontal bar shows scale. Arrows in (c) enclose a vortex ring.}
\label{fig5}
\vspace{0.0cm}
\end{figure}

To explore the parameter space, we repeated our calculations for BECs in the same harmonic trap as before, but with different values of $N$ and $n_{0}$. When we increase $N$ to $10^{6}$ ($n_{0} = 3.6 \times 10^{12}$ cm$^{-3}$), the peak in our calculated $R\left(v_{x}\right)$ curve (dot-dashed in Fig.~\ref{fig4}) occurs at a higher $v_{p}$ of $1.9$ mm s$^{-1}$ (corresponding to $\Delta x = 90$ $\mu$m). 
Simulations for BECs with a range of $N$ reveal that a plot of $\ln\left(v_{p}\right)$ versus $\ln\left(n_{0}\right)$ (Fig.~\ref{fig4} inset) is linear, with a gradient of $0.5 \pm 0.01$. This is because the peak in $R\left(v_{x}\right)$ coincides with the onset of disruption of the atom cloud so that, from inequality (\ref{eq:condition}), $v_{p} \propto \sqrt{n_{0}}\left(l_{x}/l_{r}\right) \propto \sqrt{n_{0}}$, as $\left(l_{x}/l_{r}\right)$ is independent of $N$ and $n_{0}$ for the trap considered so far. According to this analysis, for given $n_{0}$, $v_{p}$ will be higher for a more cigar-shaped BEC. To confirm this, we investigated QR for an elongated atom cloud [Fig.~\ref{fig5}(a)] formed by reducing $\omega_{x}$ to $2\pi$ rad s$^{-1}$ and increasing $N$ to $9.8 \times 10^{5}$ to maintain the original value of $n_{0}=2.2 \times 10^{12}$ cm$^{-3}$. For $v_{x}\approx2.1$ mm s$^{-1}$ ($\Delta x = 330$ $\mu$m), large side lobes form during QR [Fig.~\ref{fig5}(b)], leading to fragmentation of the atom cloud and production of a vortex ring [Fig.~\ref{fig5}(c)]~\cite{note8}. By contrast, for the same values of $n_{0}$ and $v_{x}$, the more spherical BEC shown in Fig.~\ref{fig2}(a) is not disrupted by QR [Fig.~\ref{fig3}(f)].

In summary, we have investigated the reflection of {BEC}s from three abrupt potential steps, and shown that the dynamics are qualitatively the same in each case. For high $v_{x}$, reflection occurs cleanly, causing almost no change in the atom density profile. But at low $v_{x}$, the standing wave formed from the superposition of the incident and reflected waves generates dynamical excitations that disrupt the atom cloud. This disruption could explain the anomalously low $R$ values recently measured for BECs incident on a Si surface at low velocity. Fragmentation of the atom cloud is particularly dramatic for dense slow-moving cigar-shaped BECs, and further experiments in this regime would test our interpretation of the reflection process. Conversely, for low-density {BEC}s, which are not significantly disrupted by QR, the observed values of $R$ should continue to increase with decreasing $v_{x}$, even when $v_{x}$ is small. In this regime, QR near a Si surface should provide an effective means of containing {BEC}s and decoupling them from the surface. 

We thank T. Pasquini and O. Morsch for helpful discussions. This work is funded by EPSRC UK and ARC.
\vspace{-0.5cm}
\bibliographystyle{alpha}
\bibliography{biblio}

\begin{thebibliography}{10}

\bibitem{dahan}
M.~B. Dahan {\it et~al.}, Phys. Rev. Lett. {\bf 76},  4508  (1996).

\bibitem{pethick}
C. Pethick and H. Smith, {\em Bose-Einstein condensation in dilute gases}
  (Cambridge University Press, Cambridge, 2002).

\bibitem{morsch}
O. Morsch {\it et~al.}, Phys. Rev. Lett. {\bf 87},  140402  (2001).

\bibitem{andrews}
M. Andrews {\it et~al.}, Science {\bf 275},  637  (1997).

\bibitem{mewes}
M.-O. Mewes {\it et~al.}, Phys. Rev. Lett. {\bf 78},  582  (1997).

\bibitem{chikkatur}
A. Chikkatur {\it et~al.}, Science {\bf 296},  2193  (2002).

\bibitem{bloch}
I. Bloch, T. H\mbox{\"a}nsch, and T. Esslinger, Phys. Rev. Lett. {\bf 82},
  3008  (1999).

\bibitem{myprl}
R.~G. Scott {\it et~al.}, Phys. Rev. Lett. {\bf 90},  110404  (2003).

\bibitem{mypra}
R.~G. Scott {\it et~al.}, Phys. Rev. A. {\bf 69},  033605  (2004).

\bibitem{janne}
J. Ruostekoski, B. Kneer, W. Schleich, and G. Rempe, Phys. Rev. A. {\bf 63},
  043613  (2001).

\bibitem{wu}
B. Wu and Q. Niu, Phys. Rev. A. {\bf 64},  061603  (2001).

\bibitem{fallani}
L. Fallani {\it et~al.}, Phys. Rev. Lett. {\bf 93},  140406  (2004).

\bibitem{tom}
T. Pasquini {\it et~al.}, Phys. Rev. Lett. {\bf 93},  223201  (2004).

\bibitem{shimizu}
F. Shimizu, Phys. Rev. Lett. {\bf 86},  987  (2001).

\bibitem{choi}
D.-I. Choi and Q. Niu, Phys. Rev. Lett. {\bf 82},  2022  (1999).

\bibitem{arimondo}
M. Cristiani {\it et~al.}, Optics Express {\bf 12},  4  (2004).

\bibitem{zheng}
Y. Zheng, M. Kostrun, and J. Javanainen, Phys. Rev. Lett. {\bf 93},  230401
  (2004).

\bibitem{raman}
C. Raman {\it et~al.}, Phys. Rev. Lett. {\bf 83},  2502  (1999).

\bibitem{hansel}
W. H\mbox{\"a}nsel, P. Hommelhoff, T. H\mbox{\"a}nsch, and J. Reichel, Nature
  {\bf 413},  498  (2001).

\bibitem{dutton}
Z. Dutton, M. Budde, C. Slowe, and L. Hau, Science {\bf 293},  663  (2001).

\bibitem{note9}
Inequality (\ref{eq:condition}) captures the essential physics of the
  reflection process and gives a qualitative estimate of the threshold $v_{x}$
  required to suppress disruption of the BEC\hspace{-0.13cm}  .

\bibitem{note5}
During Bragg reflection of a BEC in an optical lattice, side lobes do not form
  because the standing wave extends across the entire atom cloud. In this case,
  other mechanisms can trigger disruption of the atom
  cloud~\cite{myprl,mypra,wu,fallani,choi,arimondo}\hspace{-0.13cm}  .

\bibitem{casimir}
H. Casimir and D. Polder, Phys. Rev. {\bf 73},  360  (1948).

\bibitem{note3}
We assume that atoms which do not undergo QR are either adsorbed by the Si or
  scatter inelastically~\cite{tom}\hspace{-0.13cm}  .

\bibitem{note}
$\left|d\lambda/dx\right|$ becomes $<1$ close to the surface (for $x' \lesssim
  0.04$ $\mu$m when $v_{x} = 1$ mm s$^{-1}$). Choosing $\delta = 0.15$ $\mu$m
  ensures that $V_{\mbox{III}}\left(x\right)$ incorporates most of the
  reflecting region, whilst avoiding regions where the Casimir-Polder potential
  varies too rapidly with $x'$ to allow accurate discretization of the
  wavefunction\hspace{-0.13cm}  .

\bibitem{friedrich}
H. Friedrich, G. Jacoby, and C. Meister, Phys. Rev. A. {\bf 65},  032902
  (2002).

\bibitem{note4}
At low $v_{x}$, $R$ is slightly higher when inter-atomic interactions are
  included because the interactions increase $l_{x}$, which causes atoms near
  the leading edge of the BEC to reach the potential barrier with
  $v_{x}<\omega_{x} \Delta x$\hspace{-0.13cm}  .

\bibitem{note2}
For potential II, $R$ can be determined analytically in the absence of
  inter-atomic interactions by imposing the usual wavefunction matching
  conditions at $x=\Delta x$. The $R$ values obtained in this way for a single
  atom are within $\sim 5$\% of those determined for a BEC from our numerical
  solution of Eq. (\ref{eq:gp}): indicating that inter-atomic interactions have
  little effect on the $R$ values for potential II (as shown for potential III
  in Fig.~\ref{fig4})\hspace{-0.13cm}  .

\bibitem{morschemail}
O. Morsch, Private communication (2004)\hspace{-0.13cm}  .

\bibitem{note6}
We see qualitatively similar behavior for all $n_{c} \gtrsim 0.1$
  $n_{0}$\hspace{-0.13cm}  .

\bibitem{note8}
The reflected atom cloud [Fig.~\ref{fig5}(c)] is much smaller than the incident
  cloud [Fig.~\ref{fig5}(a)] because $R$ is only 0.14\hspace{-0.13cm}  .

\end{thebibliography}
\end{multicols}
\end{document}